\documentclass[12pt]{iopart}

\newcommand{\be}{\begin{eqnarray}}
\newcommand{\ee}{\end{eqnarray}}

\begin{document}

\title[
El-Magn around Black Holes
] 
{Electromagnetic phenomena around black holes.
}

\author{A.D. Dolgov
%Content \& Services Team
}

\address{
 Novosibirsk State University, Novosibirsk, Russia \\
 ITEP, Moscow, Russia

%IOP Publishing, Temple Circus, Temple Way, Bristol BS1 6HG, UK
}
\ead{dolgov@fe.infn.it
%submissions@iop.org
}
\vspace{10pt}
\begin{indented}
\item[]August 2019
\end{indented}

\begin{abstract}

Transition of gravitational waves, produced  in the process of coalescence of black hole binaries, 
into electromagnetic radiation in magnetic field is discussed. The magnetic field is assumed to be 
created by rotating electrically charged  black holes. The process of electric charging of black holes 
due to different mobilities of protons and electrons in the surrounding plasma is described.

\end{abstract}

\section{Generation of electromagnetic radiation by gravitational waves in magnetic field \label{s-GW-to-EM} }

The mechanism of transition of electromagnetic wave (EM) into gravitational wave (GW) in magnetic field was suggested 
in 1961 by Gerzenstein~\cite{gerz} and generalized in 1973 by Zeldovich~\cite{zeld-GW-to-EM} 
to the inverse process of transformation of gravitational waves to electromagnetic radiation in magnetic field,
Contemporary formulation of the theory was presented by Raffelt and Stodolsky~\cite{RS} in 1988.

%dsp

The transition of a plane gravitational wave, ${\sim \exp (-i \omega t + i {\bf kx} )}$, into an EM one in external
transverse magnetic field ${B_T}$ is governed by the equations:
\be
%\label{system3}
(\omega^2 - k^2) %\partial_{\mathbf{x}}^2)
h_{j} (\mathbf{k}) =\kappa k %\partial_{\mathbf{x}}
A_j(\mathbf{k})B_{T}\,,
 %\nonumber 
 \label{d2-hj}
 \\ 
{ { (\omega^2 - k^2 -  m_\gamma^2) % +i \omega \Gamma) %\partial_{\mathbf{x}}^2)
%A_j - {m^2} 
A_j (\mathbf{k})
 =\kappa k %\partial_{\mathbf{x}}
h_j(\mathbf{k})B_{T}\,,}}
%\nonumber
\label{system4}
\ee
%where $\bm{B_T}$  is the component of $\bm{B}$ orthogonal to the graviton propagation, 
where $j$ describes the polarization state
of the graviton or photon, and $h_j$ is canonically normalized field of GW, such that
the kinetic term has the form ${ (\partial_\mu h_j )^2}$, i.e. ${h_j}$ is related to the metric 
${g_{\mu\nu} = \eta_{\mu\nu}  +  \tilde h_{\mu\nu} }$ according to the relation
\be
h_j = \tilde h_j /\kappa,\,\,\,\kappa^2 = 16 \pi /m_{Pl}^2,\,\,\,m_{Pl} \approx 2\cdot 10^{19}\, {\rm GeV}.
\label{h-j}
\ee
%where $\kappa^2 = 16 \pi /m_{Pl}^2$, with $m_{Pl} \approx 2\cdot 10^{19}$ GeV is the Planck mass.

The ${m_\gamma}$-term in equation above is the effective mass of photon in the medium.  It includes the
plasma frequency ${\Omega}$ and the Heisenberg-Euler correction. Under the conditions of the problem,  
which are mostly valid in what follows,
${m_\gamma}$ is dominated by the first term,
${\Omega^2}$:
\be%\tcred{\bm{
m^2  =  \Omega^2 - \frac{2\alpha  C\omega^2 }{45 \pi}  \left( \frac{B}{B_c}\right)^2 \approx \Omega^2, 
%}}
\label{m2}
%\nonumber
\ee
where ${B_c = m_e^2/ e}$, ${e^2 = 4\pi\alpha = 4\pi/137}$. and
${C}$  is a numerical constant of  order unity. It depends upon the  relative directions of the vector $\bf B$ and the wave
polarization. The plasma frequency is equal to:
\be  
\Omega ^2 = {n_e e^2}/{m_e},
%}\nonumber
\ee
${n_e}$ is the density of electrons; while ions are neglected here.

The frequency of the gravitational waves  registered by LIGO is small in comparison with the
plasma frequency of the interstellar medium. That's why the second term in expression for ${m_\gamma}$
can be neglected. However, in the
case of larger ${\omega}$ and/or  large magnetic fields the two terms in this equation may become equal and this would lead
{to a strongly amplified resonance transition of graviton to photon} 
{(analogous to MSW resonance in neutrino oscillations)}. This case is considered at the very end of the paper.

The eigenvalues of the wave vector of the system of equations: (\ref{d2-hj}) and (\ref{system4}) are:
\be
%\tcred{\bm{
k_1 =  \pm \omega \sqrt{1 + \zeta^2} ,\,\,\,  k_2 = \pm  i m \sqrt{(1 - \zeta^2)(1-\eta^2)}, 
\label{lambda2}
%\nonumber
\ee
where
\begin{equation}
%\tcblue{\bm{
\zeta^2 = (\kappa B)^2 /m^2 \ll 1,  \,\,\,\, \eta^2 = \omega^2 / m^2.  % , \,\,\, \lambda = \bar \lambda /m .
\label{prmtrs}
%\nonumber
\end{equation}
To this eigenvalues correspond respectively the following eigenfunctions: 
\be %\tcviol{\bm{
A_1 = \eta\, \zeta h_1,\,\,\,\,
h_2 =  i\zeta  A_2 .
\label{Apm}
%\nonumber
\ee
${A_1}$ describes gravitons entering into magnetic field and creating a little photons, while ${h_2}$
describes photon screating a little gravitons. 
{In the second case the wave vector ${k_2}$  is purely 
imaginary, corresponding to damping of EM waves when ${\omega <  \Omega}$.}
{In the first case the wave vector ${ k_1}$ is real, the electromagnetic wave does not attenuate and keeps on 
running together with the gravitational wave, despite its low frequency.}
{ The gravitational wave carries the electromagnetic 
companion and does not allow it to damp, despite small ${\omega <  \Omega}$.}

\section{Heating the plasma by EM wave created by GW.} 

The interaction of electromagnetic wave with a medium is described by the dielectric permittivity ${\epsilon}$:
%which determines the relation between the wave vector and the frequency of the electromagnetic wave  
${ k^2 = \epsilon \omega^2}$. For the first solution  ${k \approx \omega}$  up to some small corrections 
of the order of  ${\zeta^2}$. {To estimate the photon loss of energy due to electron heating
we need  to know the imaginary part of  ${\epsilon}$.} According e.g. to the book~\cite{LL},
for the transverse wave this imaginary part is 
\be  %\bm{\tcred{
{\rm Im}\, \epsilon  = \sqrt{\frac{\pi}{2} }\,\frac{\Omega}{\omega k a_e} \approx \sqrt{\frac{\pi}{2}}\,\frac{\Omega}{\omega^2 a_e},
\label{Im-eps}
%\nonumber
\ee
where ${ a_e = \sqrt{T_e/ (4 \pi e^2 n_e)}}$ is the Debye screening length for electrons and ${T_e}$ is their temperature.

In the approximation of the collisionless plasma this lost energy goes  from the 
electromagnetic wave to the plasma and back. However, an
account of the electron collisions leads to the heating of the plasma by the energy of the photons which are created by the
gravitational wave.  {Hence an excessive electromagnetic  radiation with higher frequency, due to thermalization, from the 
heated plasma may be registered.}

For the interstellar medium with the electron density $n_e = 0.1\, {\rm cm}^{-3}$ and the temperature $T_e = 1 $ eV,
the Debye length is equal to 
{$a_e \approx 10^3\,\rm{cm} =3\cdot 10^{-8}$~s,}  the plasma frequency is about 
{$ \Omega \approx 3\cdot 10^4 \,{\rm sec}^{-1}$}, while the frequency of the first registered LIGO event is 
{$\omega \approx 2000/$sec.} Correspondingly $\Omega a_ e \approx 10^{-3}$ and thus
$\omega^2 {\rm Im}\,\epsilon \sim \Omega / a_e$ is much larger than $\Omega^2$. So the  amplitude of
the electromagnetic wave, carried by the gravitational wave  is given by:
\be %\tcred{\bm{
A_j \approx  \frac{\omega a_e \kappa B}{\Omega} \, h_j .
% \frac{\omega a_e}{\Omega^2 }\,   h_j  = \left( \frac{\omega}{\Omega}\right)^3 \,a_e \kappa B.
%}}\nonumber
\label{A-of-h}
\ee

Hence the energy flux of the photons absorbed by the plasma makes the following fraction of the energy flux of the
parent gravitational wave:
\be 
K \equiv \frac{L_\gamma }{L_{GW}} =\left( \frac{\omega a_e \kappa B}{\Omega} \right)^2. 
\label{fraction}
\ee 

According to the analysis of the LIGO group the total energy emitted by the gravitational waves, 
lasting approximately 0.01 seconds, is about ${3 M_\odot}$  Thus the energy flux of the gravitational waves at 
the distance $R$ from the source is  
\be 
L_{GW} \approx 100 M_\odot / (4 \pi R^2)\, {\rm per}\,\,\, {\rm second }.
\label{rho-GW}
\ee

%The frequency of these gravitons (and the produced photons) is about $\bm{10^3}$~Hz, which is by far smaller than 
%the temperature of the
%interstellar or intergalactic media and even smaller than the temperature of CMB. 
%Hence, even if the plasma absorbs
%a lot of energy 
So, ${K \ll 1}$ and the direct heating of the plasma  would not be  gigantic 
in the objects which are not too close  to the binary. On the other hand, one should remember about huge energy
carried by the gravitational waves and about weak energy absorption in the approximation of collisionless plasma.
We see now that beyond  this approximation the absorbed anergy may be much larger,.

%ndeed,
%However, this is not all the truth because
The electrons in the plasma can be accelerated by the electric field of the running electromagnetic wave and obtain a very
large energy. Indeed, the electrons in the electric field of the wave are accelerated according to the equation:
\be 
m_e \ddot x_e = e E = e E_0 \cos (\omega t)
\ee
and acquire the velocity 
\be 
V_e \sim \ddot x_e /\omega \sim e E_0 / (m_e \omega)
\label{V-e}
\ee
where $ \omega $ is of the order of the frequency of the incoming gravitational wave. 
So the electrons could gain the energy:
\be 
{\cal E}_e = \frac{m_e V_e^2 }{2} \sim \frac{e^2 E_0^2}{ m._e \omega^2} .
\label{E-e}
\ee
This result is true if the electron collision time due to Compton (Thomson) or Coulomb scattering is much longer than the
inverse frequency of the wave. This condition is normally fulfilled for the interstellar or intergalactic plasma.

If we take the distance $R$ equal to the gravitational radius of the black hole, $r_g = 2M /m_{Pl}^2$, then for 
the mass $ M= 30{M_\odot}$, we find
${R = r_g = 10^7}$ cm, and the electrons  would accelerate up to the energy  
{${ {\cal E}_e = 4 {\rm eV} (B/{\rm Gauss})^2 }$,} becoming relativistic for rather mild fields $ B \geq 10^3$ Gauss.
In such plasma $e^+e^-$ pairs must be created. Their presence would change the values of the plasma frequency and 
of the Debye length but qualitatively the picture would remain essentially the same.

{The presented above estimate is obtained under assumption of homogeneous external magnetic field, i.e. for the case
when the wave length ${\lambda}$ is much smaller than the scale of the field variation, ${l_B}$. In the opposite limit 
the effect would be suppressed by the factor ${l_B/\lambda}$}.

{If such gravitational wave heats a magnetar with magnetic field about ${10^{15}}$ G, the produced burst of 
electromagnetic 
radiation would be significant for the distances between the magnetar and the coalescing black holes up to ${10^{-5}}$ 
parsec, which is very small by the astrophysical scales and is improbable.}

\section{Magnetic field generation around rotating bodies}

{Much more powerful could be the burst of electromagnetic radiation if the black hole binary is surrounded by the medium 
with sufficiently strong magnetic field.}
{ Such a field may be created by an analogue of the  Biermann battery~\cite{BB}
induced by the rotating binary due to the different mobility of protons and electrons in the surrounding
bath of electromagnetic radiation.}

As is known that the difference between masses of proton and electron  results in 
four million times difference 
of their elastic scattering on photons and hence in a difference of 
their mobilities in interstellar plasma. {Thus it
leads to predominant capture of protons by
celestial bodies, making them electrically charged}
(Shwarzman mechanism of star charging~\cite{star-charge-1,star-charge-2}).

Accordingly the interstellar medium around a star also becomes electrically charged.
The charge of a black hole surrounded by  plasma of protons and electrons
was calculated in ref.~\cite{BDP}.
%{Bambi, AD, Petrov, JCAP 0909 (2009) 013 - charging of stellar bodies}\\
Equations governing the motion of electrons and protons in gravitational and electric field 
created by a black hole with mass $M$ and electric charge $Q$
%surrounding black hole 
in spherically symmetric case have the form: 
\be
\label{dot-vp} 
\dot v_p = -\frac{G_N M}{r^2} 
+ \frac{\alpha Q}{r^2 m_p} 
+ \frac{L \, \sigma_{\gamma p}}{4 \pi r^2 m_p} 
- \frac{\sigma_{\gamma p} n_\gamma \omega_\gamma}{m_p} \, v_p
- \frac{n_p \sigma_{pe} P}{m_p} \, (v_p - v_e),  
\\
\dot v_e = -\frac{G_N M}{r^2} 
- \frac{\alpha Q}{r^2 m_e} 
+ \frac{L \, \sigma_{\gamma e}}{4\pi r^2 m_e} 
- \frac{\sigma_{\gamma e} n_\gamma \omega_\gamma}{m_e} \, v_e
+ \frac{n_e \sigma_{pe} P}{m_e} \, (v_p-v_e) \, .
 %\nonumber
\label{dot-ve} 
\ee
Here ${v_p}$ and $v_e$ are the proton and electron {regular} velocities,
%(that is the average velocities of $p$ and $e$ in the plasma 
%which do not include the large chaotic thermal velocities), 
$Q$ is the electric charge of the BH in proton charge units, 
${\alpha = e^2/4\pi = 1/137}$, 
$\sigma_{ij}$ is the cross section of scattering of particle $i$ on particle $j$, 
$L$ is the BH luminosity in the comoving 
frame of the accretion flow, $P$ is the momentum transfer in 
$ep$--scattering, $n_p$ and $n_e$ are the number densities of 
$ p$ and ${e}$, %around the BH, 
${n_\gamma}$ is the photon number 
density %in the photon bath surrounding the BH 
and ${\omega_\gamma}$ is the photon energy.
%which is roughly the momentum transfer 
%in $p\gamma$-- and $e\gamma$--scattering. 
%We neglect the angular
%momentum term because, as explained below, we are interested in
%the particles with low angular momentum. 

%%Eqs.~(\ref{dot-vp}) and (\ref{dot-ve}) indeed recover the Eddington 
%%limit for $Q=0$ and stationary flux $\dot{v}_j=0$ if the fourth term 
%%on the right hand side of the two equations, that is, the term due to 
%%scattering on the thermal bath of photons, is neglected. 
%%Taking $Q=0$, $n_p = n_e$ and $\dot{v}_j=0$, we can solve the 
%%two equations for $L$, finding the usual Eddington luminosity, 
%%$L_E = 4 \pi G_N M m_p / \sigma_{e\gamma}$.

As is shown in ref.~\cite{BDP}, these equations have stationary solution with $Q$ tending to a constant value, when
the gravitational attraction of protons is counterweighted by the Coulomb repulsion:
\be
\alpha Q \approx \frac{r_g m_p}{4 \kappa^{1/4}} 
\left(T_e/T_p\right)^{3/4} 
\sqrt{\frac{\ln(\lambda_p/r_g)}{\ln(\lambda_e/r_g)}}
\label{alpha-Q}
\ee
where $\kappa = m_p/m_e$, $\lambda_{e}$ and $\lambda_{p}$ are the mean-free paths of $e$ and $p$ respectively,
and $T_e$ and $T_p$ are their temperatures,.

For $T_p=T_e$, we obtain $\alpha Q \approx 0.15 \, m_p r_g$, leading to  the electric field at the gravitational radius:
\be
E = \frac{\alpha Q}{r_g^2} \approx 0.15 \frac{m_p}{r_g} = 0.075 \frac{m_p}{M} \, m_{Pl}^2 
\ee  
and the ratio of the critical Schwinger field to this $E$  equal to:
\be
\frac{E_c}{E} = 0.6 \left(\frac{M}{10^{20} \; {\rm g}}\right) \, .
\ee

 Rotating locally charged sphere
creates non-zero magnetic fields.
Normally the strength of such magnetic field is given by the Biot-Savart law. 
However in astrophysical systems  the time of establishment this law may be too long~\cite{BDT}.
%Z.Berezhiani, AD, I.Tkachev, {"Dark matter and generation of galactic magnetic fields",}
%The European Physical Journal, C73 (2013) 2620.\\

The Biot-Savart law is valid only when the stationary regime is reached, 
but the system under scrutiny may be far from that. 
The time to reach the stationary situation can be much longer than the cosmological time. 
To see that let us consider the Maxwell equations in the cosmological plasma 
and modification of the MHD equations in presence of extra non-potential forces 
related to a dark matter interaction with electrons. 
Namely,  let us consider the electric current 
\be 
{\vec J} = \sigma ({\vec E} +  {\vec v} \times {\vec B} + {\vec F}/e),
\ee 
where ${\vec F}$ is the external  force acting  on electrons, 
$ {\vec F} = e {\vec v}  B_F$
%}}\nonumber
where the factor $B_F$ can be estimated as 
$
%\label{B-gamma} 
B_F =  \sigma_{e\gamma}^{~} n_\gamma \omega_\gamma/e  .$
%= 3.4 \times 10^{-30}\, (1+z)^4 ~{\rm eV}^2/e
%= 5.8 \times 10^{-28}\, (1+z)^4 ~{\rm G} \, .
%}}\nonumber
%\ee

%see Eq.~(\ref{F-e}).  
%In our case it is the drag force induced by the interaction with the CMB 
%(or with dark matter halo, see below). 

Finding electric  field ${\vec E}$ from  the equation for $J$ in the previous page and 
substituting it into equation  $\partial_t {\vec B} = - {\vec \nabla} \times {\vec E}$, 
we obtain 
%\tcmag{$\bm{\partial_t \vB = \vnabla \times (\vv \times \vB + \vF/e - \vJ / \sigma)}$. } 
%Substituting the above expression for $\vJ$  and using 
%$\vnabla \times \vE =- \partial_t \vB$ and $\vnabla \cdot \vB =0$, we come to
%
\begin{equation} \label{MHD}
 \partial_t {\vec B} =  {\vec \nabla}\! \times \! {\vec F}/e \, + \, {\vec \nabla} \! \times \! ({\vec v} \! \times \! {\vec B}) \, + \, 
 \frac{1}{4\pi\sigma} (\Delta {\vec B} +  \partial^2 _t {\vec B}),  
\end{equation} 
which is in fact the MHD equation  in  the presence of the external source
term 
\be
{\vec \nabla} \! \times \! {\vec F}/e= B_F {\vec \nabla} \! \times {\vec v} \, + \, ({\vec \nabla} B_F)\times {\vec v.} 
\ee
In the limit of high conductivity, the second term in the MHD equation, 
the so called advection term, leads to a dynamo effect on the magnetic seed fields 
once the value of the latter is non-zero. 
It is well-known, however, that in absence of the source term, 
the MHD equations cannot give rise to non-zero magnetic field if ${\vec B}=0$ initially.

In our case, assuming ${\vec B}=0$ at $t=0$, we find that the source term 
induces a nonzero magnetic seed field which initially grows 
roughly as
\be
{\vec B(t)} =  \int_{0}^t  dt \, {\vec \nabla} \! \times \! {\vec F}/e  
=  \int_{0}^t dt \,  {\vec \nabla} \! \times \! (B_F{ \vec v}) \,.
\nonumber
\ee
%
%%However, taking into account  that $B_F(t) \propto (1+z)^4 \propto t^{-8/3}$, see eq. (\ref{B-gamma}), 
%%we find that the biggest value for the magnetic seeds can be obtained around the 
%%cosmological epoch of hydrogen recombination and photon decoupling, 
%%$z  \sim 1000$, or $t  \sim 5 \times 10^5$~yr. 
%%At earlier times, the plasma is strongly coupled and the relative motion of the electrons 
%%with respect to protons is negligible, hence the lower limit of the integration is irrelevant.  
%%The mean value of vorticity at the characteristic spatial scale $\lambda$ is  
%%$\Omega_\lambda = | \vnabla \! \times \! \vv |_\lambda  \leq  10^3(\delta T/T)^2/\lambda$, 
%%as calculated in Ref. \cite{ber-dol}. 

%%}

%%\frame{
%%\frametitle {\LARGE \myfont Magnetic field generation around rotating binary}

{Then one can formally estimate that at large time the strength of magnetic field
generated near rotating binary of black holes with masses about 20-30 solar masses
to be around ${10^{10} }$ Gauss. However, this field is much larger than the Biot-Savart limit, which cannot be
violated. Thus the maximum magnetic field generated by rotating electrically charged black hole 
could be not larger than $0.1 (10^{35} {\rm g} ) / M$ Gauss. So for the LIGO events it is negligibly small but
for  much less massive BHs it can even exceed the critical value $10^{13} $ G.

\section*{References}
%\verb"

\section*{Acknowledgement} 
The work  was supported by the RSF Grant 19-42-02004. 

\end{document}